\documentclass[12pt]{article}

\usepackage{amsmath}
\usepackage{graphicx}
\usepackage{amsfonts}
\usepackage{amssymb}
\usepackage{epsfig}
\usepackage{color}

\newcommand{\resection}[1]{\setcounter{equation}{0}\section{#1}}

\newcommand{\Cblu}[1]{\textcolor{blue}{#1}}
\newcommand{\Cgre}[1]{\textcolor{green}{#1}}
\newcommand{\Cred}[1]{\textcolor{red}{#1}}

\begin{document}
\setcounter{page}{0} \topmargin 0pt \oddsidemargin 5mm
\renewcommand{\thefootnote}{\arabic{footnote}}
\newpage
\setcounter{page}{0}

\begin{titlepage}

\begin{flushright}
SISSA/54/2001/FM
\end{flushright}

\vspace{0.5cm}
\begin{center}
{\large {\bf Boundary Quantum Field Theories with }}\\
{\large {\bf Infinite Resonance States}} \\
\vspace{2cm}
{\large P. Mosconi$^{1,2}$, G. Mussardo$^{1,3}$ and V. Riva$^{1}$} \\
\vspace{0.5cm} {\em $^{1}$International School for Advanced
Studies, Trieste, Italy }\\
\vspace{0.3cm} {\em $^{2}$Istituto Nazionale di Fisica della Materia,
Sezione di Trieste}\\
\vspace{0.3cm} {\em $^{3}$Istituto Nazionale di Fisica Nucleare, Sezione di Trieste}
\end{center}
\vspace{1cm}

\begin{abstract}
We extend a recent work by Mussardo and Penati  on integrable quantum field theories with a single stable
particle and an infinite number of unstable resonance states, including the presence of a boundary. The
corresponding scattering and reflection amplitudes are expressed in terms of Jacobian elliptic functions, and
generalize the ones of the massive thermal Ising model and of the Sinh-Gordon model. In the case of the
generalized Ising model we explicitly study the ground state energy and the 
one--point function of the thermal
operator in the short--distance limit, finding an oscillating behaviour related to the fact that the infinite
series of boundary resonances does not decouple from the theory even at very short--distance scales. The analysis
of the generalized Sinh-Gordon model with boundary reveals an interesting constraint on the analytic structure of
the reflection amplitude. The roaming limit procedure which leads to the Ising model, in fact, can be
consistently performed only if we admit that the nature of the bulk spectrum uniquely fixes the one of resonance
states on the boundary.
\end{abstract}

\end{titlepage}

\newpage

\resection{Introduction}

The study of resonances in relativistic models has provided important insights into the general problem of the
dynamics of quantum field theories. Resonances are usually due to unstable bound states which therefore do not
appear as asymptotic states. However, associated to them, there are mass scales which may induce interesting and
unexpected phenomena. Al. Zamolodchikov, for instance, has shown in \cite{roaming} that a single resonance state
produces a remarkable pattern of roaming Renormalization Group trajectories, with the property that they pass by
very closely all minimal unitary models of Conformal Field Theory, finally ending in the massive phase of the
Ising model. The first model with an infinite number of resonance states is due to A. Zamolodchikov, in relation
with a QFT characterized by a dynamical $Z_4$ symmetry \cite{Z4Zam} (see also \cite{LukyZam}). A simpler class of
models with infinitely many resonance states have been considered by Mussardo and Penati in \cite{bulk}. The
analysis performed in \cite{bulk} of the Form Factors and the correlation functions of these theories has shown a
rather rich and still unexplored behaviour of such theories. A further step in the analysis of infinite resonance
models has been taken by Castro--Alvaredo and Fring in \cite{Fring}.

The aim of this paper is to extend the analysis of the physical effects induced by an infinite tower of resonance
states, this time also placed at the boundary of a $1 + 1$ dimensional quantum field theory. We will consider, in
particular, the boundary dynamics of two models which were previously defined in \cite{bulk}. Those may be
regarded as the generalizations of the massive thermal Ising model and of the Sinh-Gordon model, respectively. In
the bulk both theories have a single stable excitation of mass $\textrm{M}$. 
Our main purpose is to analyse the boundary behaviour of these theories  
and to show some novel phenomena induced by the resonance states. In 
particular, for the generalization of the Ising model we study the ground 
state energy and the one--point correlator of the thermal
operator in the short--distance limit. Contrary to the ordinary case, 
where these functions scale as a constant
and as an inverse power respectively, here they both assume an 
oscillating behaviour, hinting to the fact that
the infinite series of boundary resonances does not decouple from the 
theory even at very short distance scales.
The oscillating nature of these quantities prevents a standard 
conformal field theory interpretation of the
theory at short distances. Hence, the boundary seems to have a more 
drastic effect on the theory with respect to
the bulk, where the presence of an infinite number of resonances has the 
effect to make softer its ultraviolet properties leaving, however, the 
possibility to recover a conformal behaviour and to define conformal 
data such as the central charge or the anomalous dimensions \cite{bulk}.

The scattering and reflection amplitudes of the two theories investigated in this paper can be expressed, in the
rapidity variable $\theta$, in terms of Jacobian elliptic functions, which are periodic along both the complex
axis. The resonance states correspond to poles in the unphysical sheet of the scattering amplitude. The choice of
doubly periodic scattering amplitudes forces us to consider only theories without any genuine bound states,
otherwise the presence of a pole on the imaginary axis inside the physical strip would imply a proliferation of
infinite other poles located at the same imaginary position but at non vanishing real one, spoiling the causality
of the theory.

A complete list of properties of the Jacobian elliptic functions can be found in \cite{GRA}. Here we simply
recall that these functions, indicated by $\textrm{sn}(u)$, $\textrm{cn}(u)$ and $\textrm{dn}(u)$, depend on a
parameter $\ell$ (called modulus) which varies between 0 and 1, and have poles and zeroes located with double
periodicity on the complex plane of $u$. In our notation the rapidity variable $\theta$ will enter the argument
of these functions in the form $u=\frac{\textbf{K}}{i\pi}\,\theta$, where $\textbf{K}$ is the complete elliptic
integral of modulus $\ell$, defined as
$$
\textbf{K}(\ell)= \int_{0}^{\frac{\pi}{2}}\frac{d\alpha}{\sqrt{1-\ell^{2} \sin^{2}\alpha}}\,.
$$
We will also write many expressions in terms of the complementary modulus $\ell'=\sqrt{1-\ell^{2}}$ and of the
corresponding complete elliptic integral $\textbf{K}'(\ell)=\textbf{K}(\ell')$. In the limit $\ell\to 0$ only the
periodicity along the imaginary axis in $\theta$ survives, and we recover the ordinary scattering theories
without resonance states. The corresponding asymptotics of our functions are given by
$$
\textbf{K}\to\frac{\pi}{2}\qquad\textbf{K}'\rightarrow \infty
$$
$$
\textrm{sn}(u)\rightarrow \sin(u)\qquad\textrm{cn}(u) \rightarrow \cos(u)\qquad\textrm{dn}(u)\rightarrow 1
$$
We anticipate here that all the physical quantities we have studied share the same qualitative behaviour for
every $\ell>0$, while possible discontinous changes may occur for $\ell\to 0$.

\newpage

\resection{Integrable Boundary Quantum Field Theories}

Integrable quantum field theories with boundary ({\em i.e.} defined on half--line and with the boundary placed at
$x=0$) have been defined and analyzed in \cite{ghoszam,kobfrin}. Here, we will focus on systems with a single
bulk excitation and without any bulk or boundary bound states. In particular, we recall that the physical strip
is given by $0\leq\textrm{Im}\,\theta\leq\pi$ for the bulk scattering matrix $S$ and by
$0\leq\textrm{Im}\,\theta\leq\frac{\pi}{2}$ for the boundary reflection matrix $K$.

In our simple case, the two basic equations of unitarity and crossing, respectively, assume the form:
\begin{equation}\label{unit}
K\left(\theta\right)K\left(-\theta\right) = 1\,\,\,;
\end{equation}
\begin{equation}\label{cross}
K\left(\theta\right)K\left(\theta+i\pi\right)=S\left(2\theta\right)\,\,\,.
\end{equation}
Following \cite{bulk}, where the $S$-matrix was assumed to have both an imaginary and a real periodicity, we will
look for solutions to eq. (\ref{unit}) and (\ref{cross}) with the same property, i.e.
\begin{eqnarray}
\label{perS} S(\theta+2\pi i)=S(\theta)&\qquad;\qquad &
S(\theta+T_{S})=S(\theta) \,\,\,;\\
\label{perK} K(\theta+2\pi i)=K(\theta)&\qquad;\qquad;& 
K(\theta+T_{K})=K(\theta) \,\,\,.
\end{eqnarray}
Eq.\,(\ref{cross}) uniquely fixes the imaginary periodicity of $K$ to be 
$2\pi i$, while allows the real one to be
$T_{K}=T_{S}$ or $T_{K}=\frac{1}{2}T_{S}$, thanks to the fact that 
$S$ is evaluated at $2\theta$; this observation will be
useful in the following considerations.

\resection{The Elliptic Ising Model}

The simplest elliptic $S$-matrix is given by $S(\theta)=-1$, which satisfies eq. (\ref{perS}) with any choice of
$T_{S}$. Referring to \cite{bulk}, we interpret 
this as a particular analytic continuation of an $S$-matrix which
possesses an infinite number of poles and zeroes, in the limit in which they all cancel each other.

\begin{figure}[h]
\setlength{\unitlength}{0.0125in}
\begin{picture}(40,90)(60,420)
\put(190,480){\vector(1,0){230}} \put(300,410){\vector(0,1){160}} \put(412,560){$\theta$}
\put(407,558){\line(0,1){13}}\put(407,558){\line(1,0){13}} \Cgre{\put(235,510){\circle{8}}
\put(300,510){\circle{8}} \put(364,510){\circle{8}}} \Cred{\put(226,446){$\times$} \put(291,446){$\times$}
\put(356,446){$\times$}} \put(230,477){\line(0,1){6}} \put(360,477){\line(0,1){6}} \put(220,463){$-T$}
\put(360,463){$T$} \put(293,420){\line(1,0){6}} \put(293,540){\line(1,0){6}} \put(266,418){$-i\pi$}
\put(276,538){$i\pi$}
\end{picture}
\caption{Analytic structure of $K_{I}(\theta)$ in the fundamental domain} \label{anKIell}
 \end{figure}
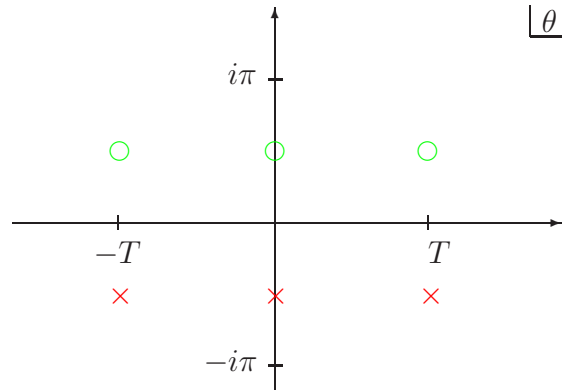

A solution to eq.\,(\ref{unit}) and (\ref{cross}) in correspondence to $S(\theta)=-1$ is given by
\begin{equation}\label{KIell}
K_{I}(\theta)=\sqrt{\ell'}\;\;\frac{\textrm{sn} \left[\frac{\textbf{K}}{i\pi}\left(\theta-i\frac{\pi}{2}
\right)\right]}{\textrm{cn}
\left[\frac{\textbf{K}}{i\pi} 
\left(\theta-i\frac{\pi }{2}\right)\right]}\,\,\,.
\end{equation}
This elliptic function has real period $T=2\pi 
\frac{\textbf{K}'} {\textbf{K}}$ with the analytic structure shown
in Figure \ref{anKIell}, where circles and crosses represent, 
respectively, the positions of zeroes and poles.

The poles, located at
\begin{equation}
\theta_{n,m}=-\,i\,\frac{\pi}{2}+i\, 2m\pi+ n T\,,
\end{equation}
correspond to an infinite set of boundary resonance states with masses and decay widths given by
\begin{eqnarray}
\label{mres}M_{\textrm{res}}(n)&=&\sqrt{2} \,\textrm{M}\;
\cosh\left(\frac{n T}{2}\right)\,,\\
\label{gres}\Gamma_{\textrm{res}}(n)&=&\sqrt{2}\, \textrm{M}\sinh\left(\frac{n T}{2}\right)\qquad n=1,2,...\,,
\end{eqnarray}
where $\textrm{M}$ is the bulk mass parameter of the theory.

\vspace{3cm}

\vspace{1cm}

In the ordinary limit $\ell\rightarrow 0 $, $K_{I}(\theta)$ in (\ref{KIell}) reduces to the amplitude
$K_{\textrm{fixed}}(\theta)=i \tanh\left(i\frac{\pi}{4}-\frac{\theta}{2}\right)$ found in \cite{ghoszam}. This is
the reflection amplitude relative to the Ising model with fixed boundary 
conditions at $x=0$. The other
ordinary solution $K_{\textrm{free}}(\theta)=
-i \coth\left(i\frac{\pi}{4}-\frac{\theta}{2}\right)$, relative to the 
Ising model with free boundary condition at $x=0$, cannot be extended to 
the elliptic case, because it would have poles with real part
in the physical strip.

The elliptic generalization of the factor interpolating between $K_{\textrm{fixed}}$ and $K_{\textrm{free}}$
given in \cite{ghoszam} is
\begin{equation}\label{CDD}
\frac{\textrm{sn}(2\textbf{K}c)-\textrm{sn}\left(\frac{i\textbf{K}}{\pi}2 \theta\right)}
{\textrm{sn}(2\textbf{K}c)+\textrm{sn}\left(\frac{i\textbf{K}}{\pi}2 \theta\right)}=
-\frac{\textrm{sn}\left[\frac{\textbf{K}}{i\pi}(\theta+i\pi
c)\right]}{\textrm{sn}\left[\frac{\textbf{K}}{i\pi}(\theta-i\pi
c)\right]}\frac{\textrm{cn}\left[\frac{\textbf{K}}{i\pi}(\theta-i\pi
c)\right]}{\textrm{cn}\left[\frac{\textbf{K}}{i\pi}(\theta+i\pi
c)\right]}\frac{\textrm{dn}\left[\frac{\textbf{K}}{i\pi}(\theta-i\pi
c)\right]}{\textrm{dn}\left[\frac{\textbf{K}}{i\pi}(\theta+i\pi c)\right]}\,,
\end{equation}
where the parameter $c$ is related to the \lq\lq boundary magnetic field \rq\rq\, $h$ by the relation
\begin{equation}
\textrm{sn}(2\textbf{K}c)=1-\frac{h^{2}}{2\,\textrm{M}}\,\,\,.
\end{equation}

\newpage

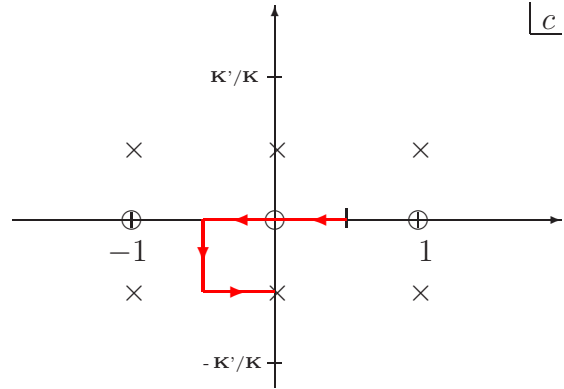
\begin{figure}[h]
\setlength{\unitlength}{0.0125in}
\begin{picture}(40,90)(60,420)
\put(190,480){\vector(1,0){230}} \put(300,410){\vector(0,1){160}} \put(412,560){$c$}
\put(407,558){\line(0,1){13}}\put(407,558){\line(1,0){13}} \put(236,506){$\times$} \put(296,506){$\times$}
\put(356,506){$\times$} \put(240,480){\circle{8}} \put(300,480){\circle{8}} \put(360,480){\circle{8}}
\put(236,446){$\times$} \put(296,446){$\times$} \put(356,446){$\times$} \put(240,477){\line(0,1){6}}
\put(360,477){\line(0,1){6}} \put(230,463){$-1$} \put(360,463){$1$} \put(297,420){\line(1,0){6}}
\put(297,540){\line(1,0){6}} \put(270,418){\tiny -\,\textbf{K}'/\textbf{K}}
\put(273,538){\tiny\textbf{K}'/\textbf{K}} \put(330,477){\line(0,1){8}} \thicklines
\Cred{\put(330,480){\vector(-1,0){15}} \put(330,480){\vector(-1,0){48}} \put(330,480){\line(-1,0){60}}
\put(270,480){\line(0,-1){30}} \put(270,480){\vector(0,-1){18}} \put(270,450){\vector(1,0){18}}
\put(270,450){\line(1,0){30}}}
\end{picture}
\caption{Path of $c$} \label{path}
 \end{figure}

In order to describe the variation of $h$ from 0 to $\infty$, the parameter $c$ has to follow the path drawn in
Figure \ref{path}, which displays the analytic structure of $\textrm{sn}(2\textbf{K}c)$.

\vspace{2.5cm}

\vspace{0.5cm}

The important qualitative difference with the ordinary case is that now we have to impose a minimum boundary
magnetic field $h_{\textrm{min}}=\sqrt{2\,\textrm{M}}$, in order to start from the point $c=0$ in the path of
Figure \ref{path}. In fact, for $c\in[0,\frac{1}{2}]$ the factor (\ref{CDD}) has poles with real part in the
physical strip, and in particular the value $c=\frac{1}{2}$ (or equivalently $h=0$) would correspond to the
ordinary reflection amplitude $K_{\textrm{free}}$. Hence, switching on the elliptic parameter $\ell$ from zero to
a finite value, the minimum admitted value of $h$ changes discontinuously from 0 to $\sqrt{2\,\textrm{M}}$.

The general reflection amplitude assumes then the form
\begin{equation}\label{KIellh}
K_{I}^{(h)}(\theta)=\sqrt{\ell'}\;\;\frac{\textrm{sn} \left[\frac{\textbf{K}}{i\pi}\left(\theta-i\frac{\pi}{2}
\right)\right]}{\textrm{cn}\left[\frac{\textbf{K}}{i\pi}
\left(\theta-i\frac{\pi}{2}\right)\right]}\cdot\frac{1-\frac{h^{2}}
{2\textrm{M}}-\textrm{sn}\left(\frac{i\textbf{K}}{\pi}2\theta\right)} {1-\frac{h^{2}}{2\textrm{M}}+\textrm{sn}
\left(\frac{i\textbf{K}}{\pi}2 \theta\right)}\,,\qquad \sqrt{2\textrm{M}}\leq h<\infty \,,
\end{equation}
and in the limit $h\to\infty$ it reduces to (\ref{KIell}).

\vspace{0.5cm}

As it was noticed, $K_{I}(\theta)$ given by (\ref{KIell}) has real period $T=2\pi
\frac{\textbf{K}'}{\textbf{K}}$, but eq.\,(\ref{cross}) with $S(\theta)=-1$ admits in principle solutions with
any periodicity. In particular, it is easy to see that the same analytic structure of (\ref{KIell}), but with
half period $T'=\pi \frac{\textbf{K}'}{\textbf{K}}$, is realized by
\begin{equation}\label{KI'ell}
K_{I}'(\theta)=\ell'\;\;\frac{\textrm{sn}\left[
\frac{\textbf{K}}{i\pi} \left(\theta-i\frac{\pi}{2}
\right)\right]}{\textrm{cn}\left[\frac{\textbf{K}}{i\pi} 
\left(\theta-i\frac{\pi}{2}\right)\right]\textrm{dn}
\left[\frac{\textbf{K}}{i\pi}\left(\theta-i\frac{\pi }{2}
\right)\right]}\,\,\,.
\end{equation}
Since it will be used in the following, we also note that the function
\begin{equation}\label{KI''ell}
K_{I}''(\theta)=- i \, K_{I}'(\theta)K_{I}'\left(\theta-\frac{T'}{2}\right)
\end{equation}
has again the same structure, but with real period 
$T''=\frac{\pi}{2} \frac{\textbf{K}'}{\textbf{K}}$ .

It is worth to stress, however, that the various choices for the period of the reflection amplitude have a
different physical meaning, corresponding to theories with distinct spectra of boundary excitations (see
eq.\,(\ref{mres}) and (\ref{gres})).

\subsection{Ground State Energy}

In \cite{BTBA} it was developed the technique to compute the ground state energy in integrable systems with
boundary. In particular, in the case with $S(\theta)=-1$ the resulting expression for the effective central
charge with equal boundary conditions at both sides of the strip takes the simple form:
\begin{equation}\label{TBA}
c_{\textrm{eff}}=\frac{12}{\pi^{2}}\,\lim_{r\rightarrow 0}\, r\int_{0}^{\infty}d\theta \cosh\theta
\ln\left(1+|\hat{K}(\theta)|^{2}\, e^{-2r \cosh\theta}\right)\, ,
\end{equation}
where $\hat{K}(\theta) = K\left(\frac{i\pi}{2} - \theta\right)$ is the amplitude relative to the boundary
condition under consideration. We have studied the behaviour of this quantity choosing the elliptic reflection
amplitude\footnote{Alternatively, we could have chosen $K'_{I}(\theta)$ or $K''_{I}(\theta)$ given, respectively,
by (\ref{KI'ell}) and (\ref{KI''ell}), but for convenience, we have preferred to use the simplest expression.}
$K_{I}(\theta)$ in (\ref{KIell}).

Performing the change of variable $x=r\cosh\theta$ in order to calculate the integral in (\ref{TBA}), it is easy
to see that the limit $r\rightarrow 0$ implies an evaluation of $|\hat{K}_{I}(\theta)|^{2}$ at
$\theta\rightarrow\infty$. In the ordinary case this simply corresponds to
$|\hat{K}_{\textrm{fixed}}(\theta)|^{2}\rightarrow 1$, but in the elliptic case $|\hat{K}_{I}(\theta)|^{2}$ is a
periodic oscillating function also for real values of $\theta$, and this has the consequence that the limit
procedure does not converge to a fixed value of $c_{\textrm{eff}}$. In particular, if we evaluate the limit
$r\rightarrow 0$ on the sequence $\left\{r_{n} \right\}_{n=1,2,...}$, with $r_{n}=\frac{a} {\cosh\left(\frac{n
T}{2}\right)}$, the corresponding sequence of integrals converges very rapidly to some fixed value which,
however, depends on the choice of $a$. This indicates that the scaling function $c_{\textrm{eff}}(\textrm{M} r)$,
which in the ordinary case tends to a constant as $\textrm{M} r\rightarrow 0$, keeps a dependence on the variable
$\textrm{M} r$, assuming the same values for constant values of $M_{\textrm{res}}(n)\, r_{n}$. This situation has
a clear physical interpretation in terms of the presence of boundary resonance states with infinitely increasing
energies, which do not decouple from the theory even at very high energies.

\begin{figure}[h]
\setlength{\unitlength}{0.0125in} \hspace{4.5cm}  \psfig{figure=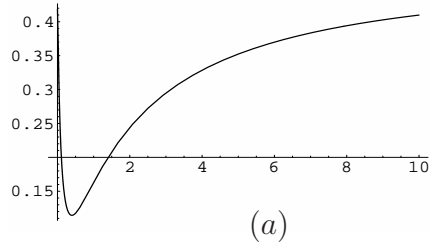,height=4cm,width=7cm} \vspace{-0.5cm}

\begin{picture}(40,90)(60,420)
\put(300,520){$(a)$}
\end{picture}

\vspace{-2.5cm}

\hspace{4.5cm} \psfig{figure=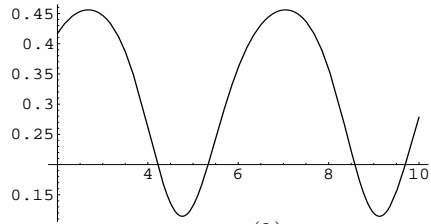,height=4cm,width=7cm} \vspace{-1cm}

\begin{picture}(40,90)(60,420)
\put(300,500){$(b)$}
\end{picture}

\vspace{-2.5cm}

\caption{$\ell=0.32$. $(a)$ $c_{\textrm{eff}}$ vs $a$, with $a\in[0,10]$, $(b)$ $c_{\textrm{eff}}$ vs $w$, with
$a=10^{-w}$ and $w\in[2,10]$.} \label{c01}
\end{figure}

\begin{figure}[h]
\setlength{\unitlength}{0.0125in} \hspace{4.5cm} \psfig{figure=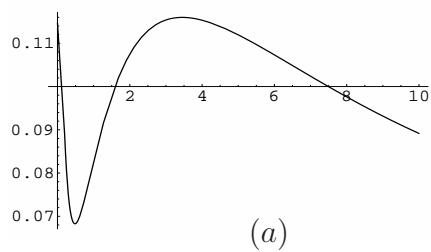,height=4cm,width=7cm} \vspace{-0.5cm}

\begin{picture}(40,90)(60,420)
\put(300,520){$(a)$}
\end{picture}

\vspace{-2.5cm}

\hspace{4.5cm} \psfig{figure=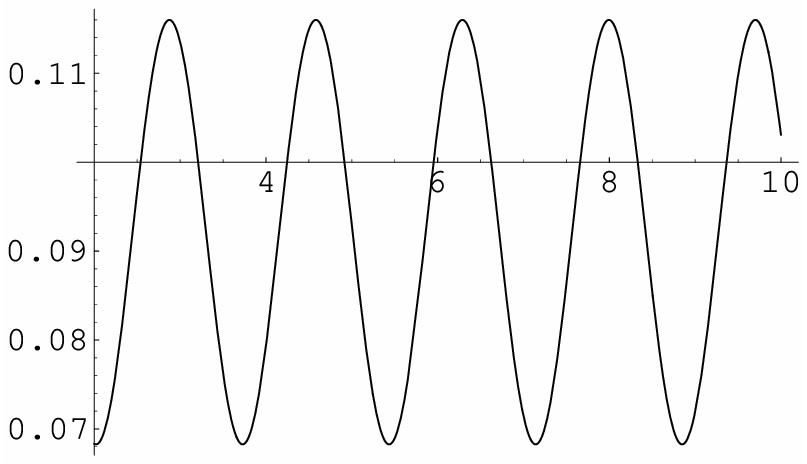,height=4cm,width=7cm} \vspace{-1cm}

\begin{picture}(40,90)(60,420)
\put(300,500){$(b)$}
\end{picture}

\vspace{-2.5cm}

\caption{$\ell=0.95$. $(a)$ $c_{\textrm{eff}}$ vs $a$, with $a\in[0,10]$, $(b)$ $c_{\textrm{eff}}$ vs $w$, with
$a=10^{-w}$ and $w\in[2,10]$.} \label{c09}
\end{figure}

Figures \ref{c01} and \ref{c09} show the behaviour of $c_{\textrm{eff}}$ 
as a function of the parameter $a$ for two values of $l$, i.e. 
$\ell=0.32$ and $\ell=0.95$, respectively. In both cases, the first plot corresponds to a variation of $a$ from 0
to 10, while the second one describes the variation of $w$ between 2 and 10, where we have defined $a=10^{-w}$.
The logarithmic scale clearly displays the oscillating behaviour, which 
leads to the absence of a given limit.

\newpage

\vspace{2cm}

\vspace{0.5cm}

\newpage

\subsection{One-point Function of the Energy Operator}

It is equally interesting to analyse the behaviour of the correlators in the presence of an infinite number of
boundary resonance states. Here we will consider the simplest correlation function.

The expression for the one-point function of the energy operator in the thermal Ising model with boundary has
been studied in \cite{BFF}, and it is given by
\begin{equation}\label{e0}
\epsilon_{0}(t)=\sum_{n=0}^{\infty}{<0\,|\,\epsilon(x,t)\,\vert\, n> <n\,\vert\,B>}\,,
\end{equation}
where $\vert\, n>$ is the asymptotic bulk state with $n$ particles, and the boundary state $\vert\, B>$ is given
by a superposition of two-particle states in which every couple consists of excitations with opposite rapidities:
\begin{equation}\label{Bstate}
\vert\, B>=\exp\left[\int_{0}^{\infty}\frac{d\theta}{2\pi} \hat{K}\left(\theta\right)
A^{\dag}(-\theta)A^{\dag}(\theta)\right]\vert\,0>\,.
\end{equation}
Using the corresponding elliptic form factor calculated in \cite{bulk}, we get the following expression for the
matrix element of the energy operator on the two-particle state, which is the only non vanishing contribution:
\begin{eqnarray}
\label{ffell} <0\,\vert\, \epsilon(x,t) \,\vert\, \theta_{1},\theta_2>&=&
 2 \pi \textrm{M} \,\ell'\,\frac{\textrm{sn}\left[\frac{\textbf{K}}
{i\pi}(\theta_1-\theta_2) \right]}{\textrm{dn}\left[\frac{\textbf{K}}
{i\pi}(\theta_1-\theta_2) \right]}\times \\
&&\exp{\left[ -\textrm{M}t(\cosh{\theta_1}+\cosh{\theta_2})+ i\textrm{M}x (\sinh{\theta_1}+\sinh{\theta_2})
\right]}\,.\nonumber
\end{eqnarray}
The correlator assumes then the form
\begin{equation}
\label{e0ell} \epsilon_0(t)=\ell'\textrm{M}\int_{0}^{\infty} d\theta\,
\frac{\text{sn}\left(\frac{\textbf{K}}{i\pi}2 \theta \right)} {\text{dn}\left(\frac{\textbf{K}}{i\pi}2 \theta
\right)} \hat{K}\left(\theta\right) e^{-2 \textrm{M}t \cosh{\theta}}\,.
\end{equation}
As before, we choose to perform the calculation with the reflection amplitude $K_{I}(\theta)$ given by
(\ref{KIell}); hence, eq. (\ref{e0ell}) specializes to
\begin{equation} \label{e0ellfixed}
\epsilon_{0}(t)=-\textrm{M} (\ell')^{3/2} 
\int_{0}^{\infty} d\theta\, E(\theta)\,e^{-2 \textrm{M} t \cosh
\theta}\,\,\,,
\end{equation}
where
\begin{equation}
E(\theta)=\frac{\textrm{sn}\left(\frac{\textbf{K}}{i\pi}2\theta \right)}
{\textrm{dn}\left(\frac{\textbf{K}}{i\pi}2\theta \right)} \cdot
\frac{\textrm{sn}
\left(\frac{\textbf{K}}{i\pi}\theta \right)} 
{\textrm{cn}\left(\frac{\textbf{K}}{i\pi}\theta
\right)}\,\,\,.
\end{equation}
We are interested to study (\ref{e0ellfixed}) in the short-distance limit. Contrary to the ordinary case
($\ell=0$), where this one-point function scales as a power law \cite{BFF}, now we expect at most a logarithmic
divergence, due to the fact that $E(\theta)$, being a periodic function without poles on the real axis, is
limited on the range of integration.

\newpage

\begin{figure}[h]
\hspace{4.5cm} \psfig{figure=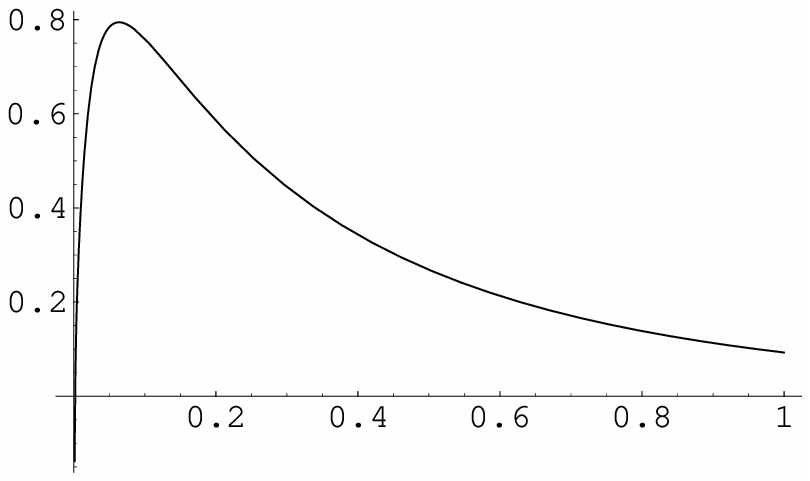,height=5cm,width=7.5cm} \vspace{-1cm} \caption{$\epsilon_{0}(r)$ for
$r\in[0,1]$\quad ($r=2\textrm{M}t$), \quad $\ell=0.71$} \label{lineare}
\end{figure}

\begin{figure}[h]
\hspace{4.5cm} \psfig{figure=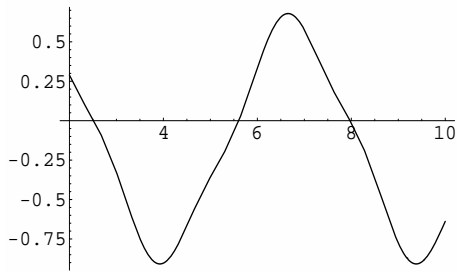,height=5cm,width=7.5cm} \vspace{-1cm} \caption{$\epsilon_{0}(w)$ with
$r=10^{-w}$, $w\in[2,10]$, \quad $\ell=0.71$} \label{log}
\end{figure}

However, as it is shown in Figures \ref{lineare} and \ref{log} (with $\ell=0.71$), the correlator neither
diverges nor goes to a finite limit as $\textrm{M}t\to 0$, but it has an 
oscillating behaviour, more visible on the
logarithmic scale.

\vspace{2cm}

\vspace{0.5cm}

\vspace{1.5cm}

\begin{figure}[h]
\hspace{3.5cm} \psfig{figure=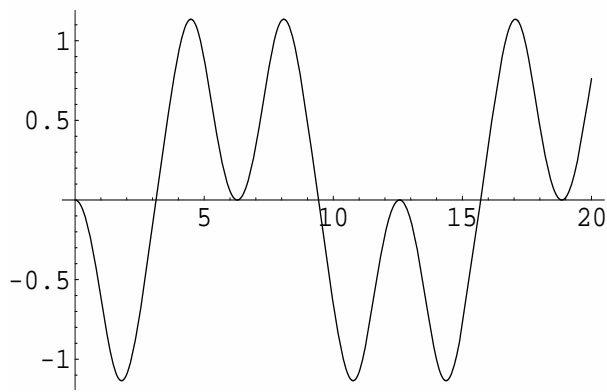,height=7cm,width=10cm} \vspace{-1.5cm} \caption{$E(\theta)$ with
$\ell=0.71$} \label{integranda}
\end{figure}

This can be easily understood analyzing the interplay between the two different factors present in the integrand.
In fact, $E(\theta)$ oscillates assuming positive and negative values with the same amplitude (Figure
\ref{integranda}), while for small values of $r$ the exponential $\exp(-r\cosh \theta)$ is approximately constant
on an interval in $\theta$ of order $\ln\frac{1}{r}$, and then falls rapidly to zero. Hence, if we decompose the
length of this interval as $\ln\frac{1}{r}=nT+x$, the integration on $nT$ gives roughly zero, while the remaining
part corresponds to an integration of $E(\theta)$ on a certain fraction of its period, so that it oscillates
depending on the value of $x$.

\vspace{0.5cm}

\vspace{2cm}

\begin{figure}[h]

\begin{picture}(40,90)(60,420)
\put(125,515){$(a)$}
\end{picture}

\vspace{-4cm}

\hspace{4cm} \psfig{figure=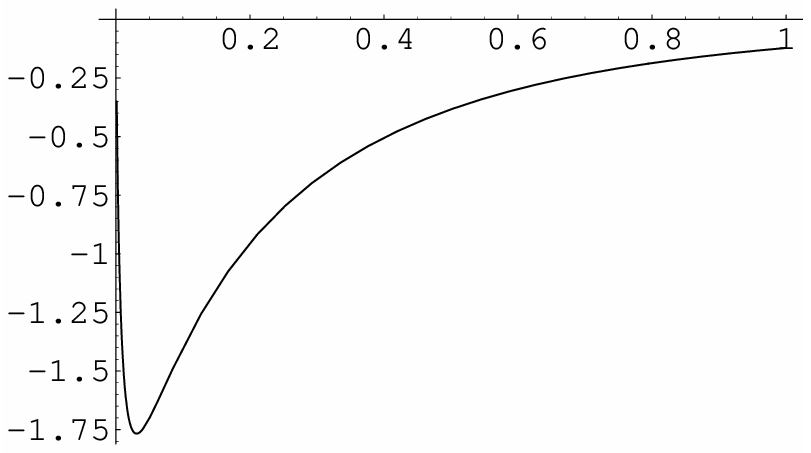,height=5cm,width=8cm} \vspace{-0.9cm}

\vspace{1cm}

\begin{picture}(40,90)(60,420)
\put(125,515){$(b)$}
\end{picture}

\vspace{-4cm}

\hspace{3.7cm} \psfig{figure=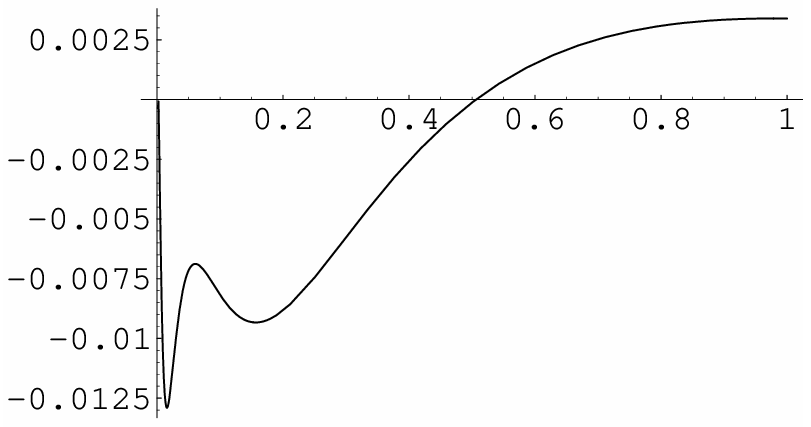,height=5cm,width=8cm} \vspace{-0.9cm}

\vspace{1cm}

\begin{picture}(40,90)(60,420)
\put(125,515){$(c)$}
\end{picture}

\vspace{-4cm}

\hspace{4cm} \psfig{figure=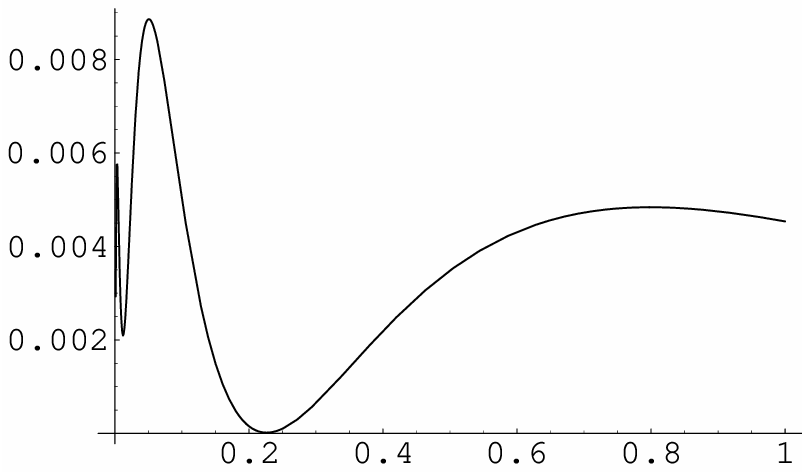,height=5cm,width=8cm} \vspace{-0.9cm}

\vspace{1cm}

\begin{picture}(40,90)(60,420)
\put(125,515){$(d)$}
\end{picture}

\vspace{-4cm}

\hspace{4.2cm} \psfig{figure=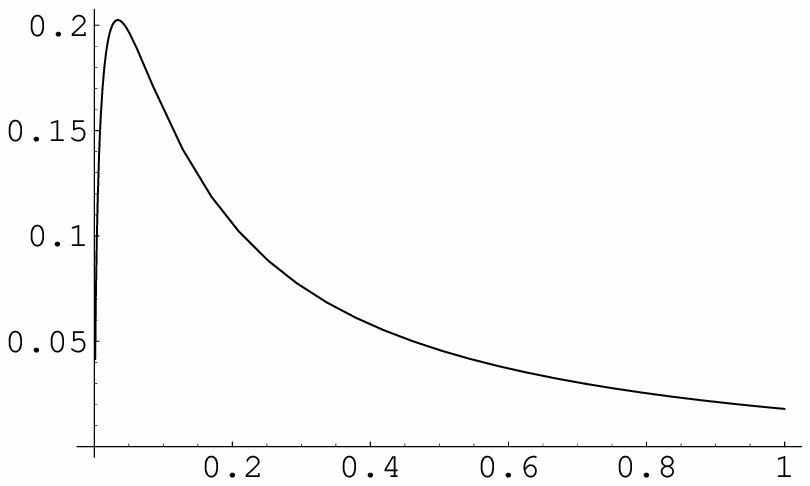,height=5cm,width=8cm}  \caption{$\ell=0.55$. One-point function of the energy
operator with $(a)$ $h=h_{\textrm{min}}$, $(b)$ $h=1.60\,h_{\textrm{min}}$, $(c)$ $h=1.61\,h_{\textrm{min}}$ and
$(d)$ $h= 1.73\,h_{\textrm{min}}$.} \label{h}
\end{figure}

Let us now turn to the reflection amplitude (\ref{KIellh}) with the boundary parameter $h$. In this case the
one-point function is given by (\ref{e0ellfixed}) with
\begin{equation}
E(\theta)\;\longrightarrow\;\frac{1-\frac{h^{2}}{2\textrm{M}}
-\textrm{sn}\left[\frac{2i\textbf{K}}{\pi}\left(i\frac{\pi}{2}-\theta\right) \right]}
{1-\frac{h^{2}}{2\textrm{M}}+\textrm{sn} \left[\frac{2i\textbf{K}}{\pi}
\left(i\frac{\pi}{2}-\theta\right)\right]}\,E(\theta)\,.
\end{equation}
For every $h_{\textrm{min}}\leq h<\infty$, at short distances these correlators do not tend to a given limit, as
before. Figure \ref{h} shows the one-point correlator as a function of $r$ for different choices of $h$, in the
case $\ell=0.55$. We note that the value $h=h_{\textrm{min}}$ corresponds just to a global change of sign with
respect to case of $K_{I}(\theta)$ (equivalent to $h\to\infty$).

Summarizing, the common behaviour shared by all these correlators 
is not surprising in the light of the previous
result about the dependence of the central charge on the 
scaling variable $\textrm{M} r$, even at short
distances. Also in this case, the absence of a proper 
ultraviolet limit can be related to the fact that the
infinite boundary resonance states never decouple from the theory. 
By using the Form Factors of the magnetic operator computed in 
\cite{bulk}, it is also easy to see that the same phenomenon 
occurs for its one--point function of this field. 

\newpage

\newpage

\resection{The Sinh-Gordon Model}

\subsection{The Ordinary Case}

The Sinh-Gordon model is defined by the field equation
\begin{equation}
\partial^{2}_{t}\phi -\partial^{2}_{x}\phi +
\frac{\sqrt{8}m^{2}}{\beta}\sinh(\sqrt{2}\beta\phi)=0\,\,\,,
\end{equation}
and has a bulk $S$-matrix of the form (\cite{arin})
\begin{equation}\label{SShG}
S(\theta)=-\frac{1}{s_{a}(\theta)s_{1-a}(\theta)}\,\,\,,
\end{equation}
where
\begin{equation}
s_{x}(\theta)=\frac{\sinh\left[\frac{1}{2}(\theta+i\pi x)\right]} 
{\sinh\left[\frac{1}{2}(\theta-i\pi x)\right]} \,\,\, ,
\end{equation}
and
\begin{equation}
a=\frac{\beta^{2}}{4\pi}\left(\frac{1}{1+\beta^{2}/4\pi}\right)\,\,\,.
\end{equation}

In the presence of a boundary, this model is integrable if the boundary conditions have the form (\cite{ghoszam},
\cite{mac})
\begin{equation}
\partial_{x}\phi|_{0}=\frac{\sqrt{2}m}{\beta}
\left(\varepsilon_{0}e^{-\frac{\beta}{\sqrt{2}}\phi(0,t)}-
\varepsilon_{1}e^{\frac{\beta}{\sqrt{2}}\phi(0,t)}\right)\,,
\end{equation}
where $\varepsilon_{0}$ and $\varepsilon_{1}$ are two boundary parameters, which can be alternatively
parameterized by
\begin{equation}
\varepsilon_{i}=\cos\pi a_{i}\qquad\qquad 0\leq a_{1}\leq a_{0}
\leq 1\,\,\,.
\end{equation}

The reflection amplitude can be obtained by analytic continuation from the one of the  sine-Gordon breather
calculated in \cite{ghos}, and has been analyzed in \cite{corrtao}. It is given by
\begin{equation}\label{Ksinh}
K(\theta)=\frac{s_{\frac{1}{2}}(\theta)s_{\frac{1+a}{2}}(\theta) s_{\frac{2-a}{2}}(\theta)}
{s_{\frac{1-E}{2}}(\theta)s_{\frac{1+E}{2}}(\theta) 
s_{\frac{1-F}{2}}(\theta)s_{\frac{1+F}{2}}(\theta)}\,\,\,,
\end{equation}
where
\begin{equation}
E=(a_{0}+a_{1})(1-a)\qquad;\qquad F=(a_{0}-a_{1})(1-a)\,\,\,.
\end{equation}
In the case of Neumann boundary conditions, where $a_{0}=a_{1}= 
\frac{1}{2}$, eq.\,(\ref{Ksinh}) becomes
\begin{equation}\label{KsinhN}
K(\theta)=\frac{s_{\frac{1+a}{2}}(\theta)} {s_{\frac{1}{2}}
(\theta)s_{\frac{a}{2}}(\theta)}\,\,\,.
\end{equation}

\subsection{The Elliptic Case}

Let us consider the elliptic $S$-matrix\footnote{In \cite{bulk} 
it was constructed and 
studied an elliptic $S$-matrix with a very
similar structure, but with half real period 
$T'=\pi \frac{\textbf{K}'}{\textbf{K}}$. Our present choice is
simply for convenient reasons. In fact, by fixing the imaginary 
period of $S$ and $K$ to be $2\pi i$, the only
two real periods which can be easily obtained multiplying 
Jacobian elliptic functions are exactly $T$ and $T'$; a
period $T''$, for example, can be obtained only in the 
more subtle way related to (\ref{KI''ell}). As we will see
below, the $K(\theta)$ is forced to have half the real period 
of the $S$--matrix. Hence, starting with the $S$
matrix of \cite{bulk}, we would have had from the beginning 
a cumbersome expression for the reflection amplitude,
which we prefer to avoid.} 
\begin{equation}\label{SsGell}
S(\theta,a)=\frac{\textrm{sn}\left[\frac{\textbf{K}}{i\pi} (\theta-i\pi
a)\right]}{\textrm{sn}\left[\frac{\textbf{K}}{i\pi} (\theta+i\pi
a)\right]}\frac{\textrm{cn}\left[\frac{\textbf{K}} {i\pi}(\theta+i\pi a)
\right]}{\textrm{cn}
\left[\frac{\textbf{K}}{i\pi}(\theta-i\pi a)\right]}\,\,\,,
\end{equation}
which has real period
$T=2\pi \frac{\textbf{K}'}{\textbf{K}}$ and in the ordinary limit ($\ell\to 0$)
reduces to the Sinh-Gordon $S$-matrix (\ref{SShG}).

\begin{figure}[h]
\setlength{\unitlength}{0.0125in}
\begin{picture}(40,90)(60,420)
\put(190,500){\vector(1,0){230}} \put(300,410){\vector(0,1){200}} \put(412,590){$\theta$}
\put(407,588){\line(0,1){13}}\put(407,588){\line(1,0){13}} \put(230,580){\line(1,0){5}}
\put(240,580){\line(1,0){5}}\put(250,580){\line(1,0){5}} \put(260,580){\line(1,0){5}} \put(270,580){\line(1,0){5}}
\put(280,580){\line(1,0){5}} \put(290,580){\line(1,0){5}} \put(300,580){\line(1,0){5}}
\put(310,580){\line(1,0){5}} \put(320,580){\line(1,0){5}} \put(330,580){\line(1,0){5}}
\put(340,580){\line(1,0){5}} \put(350,580){\line(1,0){5}} \put(360,580){\line(1,0){5}}
\Cgre{\put(234,557){\circle{8}} \put(300,557){\circle{8}} \put(364,557){\circle{8}}} \put(230,540){\line(1,0){5}}
\put(240,540){\line(1,0){5}}\put(250,540){\line(1,0){5}} \put(260,540){\line(1,0){5}} \put(270,540){\line(1,0){5}}
\put(280,540){\line(1,0){5}} \put(290,540){\line(1,0){5}} \put(300,540){\line(1,0){5}}
\put(310,540){\line(1,0){5}} \put(320,540){\line(1,0){5}} \put(330,540){\line(1,0){5}}
\put(340,540){\line(1,0){5}} \put(350,540){\line(1,0){5}} \put(360,540){\line(1,0){5}}
\Cgre{\put(234,523){\circle{8}} \put(300,523){\circle{8}} \put(364,523){\circle{8}}}
\Cred{\put(226,473){$\times$} \put(291,473){$\times$} \put(356,473){$\times$}} \put(230,460){\line(1,0){5}}
\put(240,460){\line(1,0){5}}\put(250,460){\line(1,0){5}} \put(260,460){\line(1,0){5}} \put(270,460){\line(1,0){5}}
\put(280,460){\line(1,0){5}} \put(290,460){\line(1,0){5}} \put(300,460){\line(1,0){5}}
\put(310,460){\line(1,0){5}} \put(320,460){\line(1,0){5}} \put(330,460){\line(1,0){5}}
\put(340,460){\line(1,0){5}} \put(350,460){\line(1,0){5}} \put(360,460){\line(1,0){5}}
\Cred{\put(226,439){$\times$} \put(291,439){$\times$} \put(356,439){$\times$}} \put(230,420){\line(1,0){5}}
\put(240,420){\line(1,0){5}}\put(250,420){\line(1,0){5}} \put(260,420){\line(1,0){5}} \put(270,420){\line(1,0){5}}
\put(280,420){\line(1,0){5}} \put(290,420){\line(1,0){5}} \put(300,420){\line(1,0){5}}
\put(310,420){\line(1,0){5}} \put(320,420){\line(1,0){5}} \put(330,420){\line(1,0){5}}
\put(340,420){\line(1,0){5}} \put(350,420){\line(1,0){5}} \put(360,420){\line(1,0){5}}
\put(230,497){\line(0,1){6}} \put(360,497){\line(0,1){6}} \put(213,487){\small -T} \put(365,487){\small T}
\put(273,408){$-i\pi$} \put(283,585){$i\pi$} \put(380,553){$\cdot$} \put(380,520){$\cdot$}
\put(385,553){$i\pi(1-a)$} \put(385,520){$i\pi a$} \put(237,561){\vector(4,3){23}} \put(307,561){\vector(4,3){23}}
\put(283,518){\vector(-4,-3){23}} \put(353,518){\vector(-4,-3){23}} \put(237,482){\vector(4,3){23}}
\put(307,482){\vector(4,3){23}} \put(283,439){\vector(-4,-3){23}} \put(353,439){\vector(-4,-3){23}}
\end{picture}
\caption{Analytic structure of $S(\theta)$ in the fundamental domain} \label{analS}
 \end{figure}
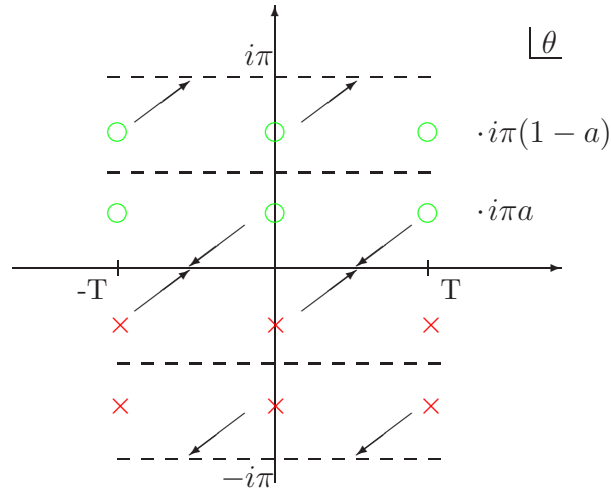

The analytic structure of $S(\theta,a)$ in shown in Figure \ref{analS}.

 \vspace{3.5cm}

\vspace{0.5cm}

In \cite{bulk} it was analyzed the generalization of the roaming limit, which leads to the $S$-matrix of the
Ising model $S(\theta)=-1$. In particular, it consists in taking
\begin{equation}
a=m+i(2n+1)\frac{T}{2\pi}+\delta \qquad m\in\mathbb{Z},\,n\in\mathbb{Z}
\end{equation}
and sending $\delta\rightarrow 0$. The arrows in Figure \ref{analS} refer to the case $m=n=0$ and indicate the
final positions of poles and zeroes, which cancel each others.

\vspace{0.5cm}

We present now an elliptic reflection amplitude with real period $T'=\pi \frac{\textbf{K}'}{\textbf{K}}$ and
without poles having a real part in the physical strip, which in the 
ordinary limit tends to the Neumann reflection
amplitude of the Sinh-Gordon model (\ref{KsinhN}):
\begin{equation}\label{KsGellN}
K(\theta)=K_{I}''(\theta)f(\theta)N(\theta)
\end{equation}
where $K_{I}''(\theta)$ is given by (\ref{KI''ell}), the factor
\begin{equation}
N(\theta)= \quad\frac{\textrm{sn}\left[\frac{\textbf{K}}{i\pi} \left(\theta-i\frac{\pi}{2}
a\right)\right]}{\textrm{sn}\left[\frac{\textbf{K}}{i\pi} \left(\theta+i\frac{\pi}{2}
a\right)\right]}\frac{\textrm{sn}\left[\frac{\textbf{K}}{i\pi} \left(\theta+i\frac{\pi}{2}
(1+a)\right)\right]}{\textrm{sn}\left[\frac{\textbf{K}}{i\pi} 
\left(\theta-i\frac{\pi}{2} (1+a)\right)\right]}\,\,\,,
\end{equation}
is the direct generalization of the ordinary one, and
\begin{equation}
f(\theta)=\left\{\frac{ \vartheta_{4}\left[\frac{1}{2 i} \left(\theta-i\frac{\pi}{2}a\right)\right]
\vartheta_{4}\left[\frac{1}{2 i}\left(\theta+i\frac{\pi}{2}(1+a) \right)\right]} {\vartheta_{4}\left[\frac{1}{2 i}
\left(\theta+i\frac{\pi}{2}a\right)\right] \vartheta_{4} \left[\frac{1}{2
i}\left(\theta-i\frac{\pi}{2}(1+a)\right)\right]} \right\}^{2}\,\,\,,
\end{equation}
is a doubly periodic function necessary to satisfy the crossing equation (\ref{cross}), which tends to $1$ as
$\ell\rightarrow 0$ (for the properties of Theta functions, see \cite{GRA}).

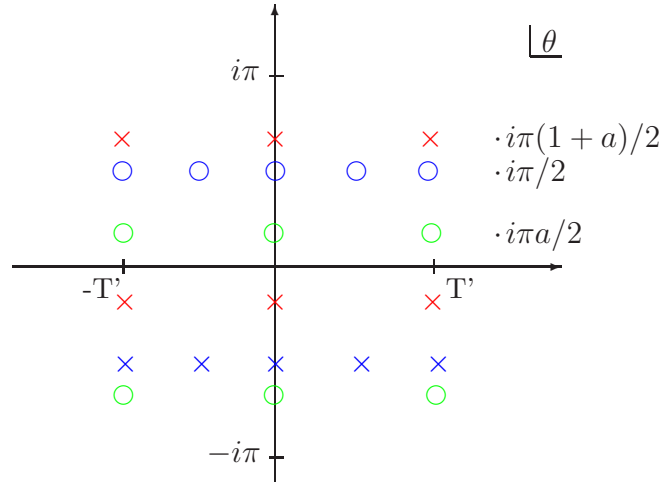
\begin{figure}[h]
\setlength{\unitlength}{0.0125in}
\begin{picture}(40,90)(60,420)
\put(190,500){\vector(1,0){230}} \put(300,410){\vector(0,1){200}} \put(412,590){$\theta$}
\put(407,588){\line(0,1){13}}\put(407,588){\line(1,0){13}} \Cred{\put(231,550){$\times$}  \put(295,550){$\times$}
\put(360,550){$\times$}} \Cblu{\put(232,540){\circle{8}} \put(264,540){\circle{8}} \put(296,540){\circle{8}}
\put(330,540){\circle{8}} \put(360,540){\circle{8}}} \Cgre{\put(228,514){\circle{8}}  \put(291,514){\circle{8}}
\put(357,514){\circle{8}}} \Cred{\put(219,482){$\times$} \put(282,482){$\times$} \put(348,482){$\times$}}
\Cblu{\put(215,456){$\times$} \put(247,456){$\times$} \put(278,456){$\times$} \put(314,456){$\times$}
\put(346,456){$\times$} } \Cgre{\put(215,446){\circle{8}} \put(278,446){\circle{8}} \put(346,446){\circle{8}}}
\put(215,497){\line(0,1){6}} \put(345,497){\line(0,1){6}} \put(198,487){\small -T'} \put(350,487){\small T'}
\put(276,420){\line(1,0){6}} \put(276,580){\line(1,0){6}} \put(250,418){$-i\pi$} \put(260,578){$i\pi$}
\put(370,536){$\cdot$} \put(370,550){$\cdot$} \put(370,510){$\cdot$} \put(375,536){$i\pi/2$}
\put(375,550){$i\pi(1+a)/2$} \put(375,510){$i\pi a/2$}
\end{picture}
\caption{Analytic structure of $K(\theta)$ in the fundamental domain} \label{analKN}
 \end{figure}

The analytic structure of $K(\theta)$ is shown in Figure \ref{analKN}

 \vspace{4cm}

\vspace{1cm}

It is interesting to note that the initial Ising factor present in 
(\ref{KsGellN}) is the one with period $T''=\frac{\pi}{2} 
\frac{\textbf{K}'}{\textbf{K}}$, while (\ref{KsGellN}) has 
altogether period $T'=\pi \frac{\textbf{K}'}{\textbf{K}}$. This choice, 
which may not seem very natural, is motivated by the requirement
that the roaming limit applied on (\ref{KsGellN}) produces the 
reflection amplitude of the Ising model.

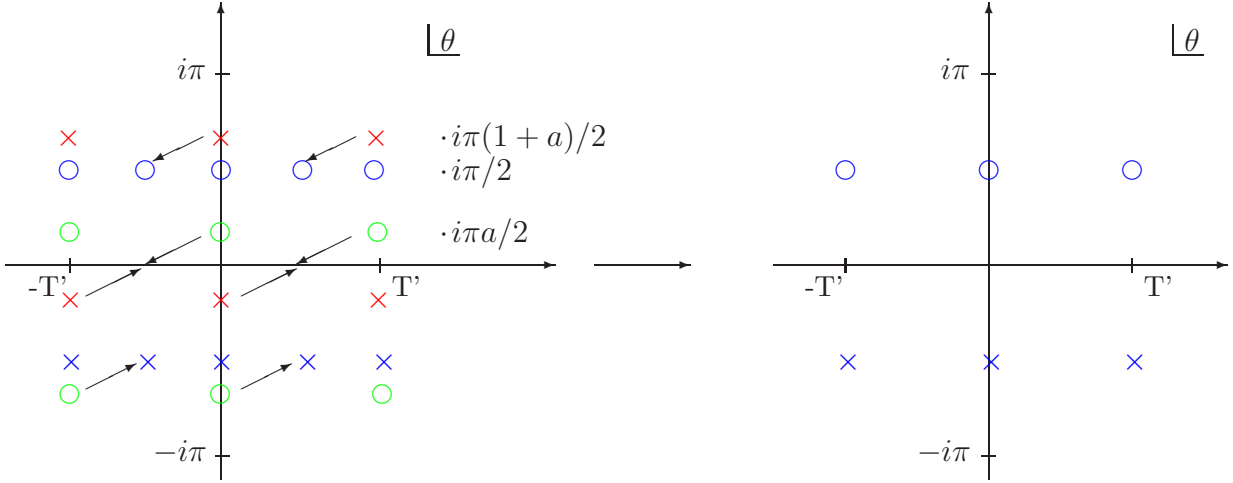
\begin{figure}[h]
\setlength{\unitlength}{0.0125in}
\begin{picture}(40,90)(60,420)
\put(50,500){\vector(1,0){230}} \put(140,410){\vector(0,1){200}} \put(232,590){$\theta$}
\put(227,588){\line(0,1){13}}\put(227,588){\line(1,0){13}} \Cred{\put(71,550){$\times$}  \put(135,550){$\times$}
\put(200,550){$\times$}} \Cblu{\put(72,540){\circle{8}} \put(104,540){\circle{8}} \put(136,540){\circle{8}}
\put(170,540){\circle{8}} \put(200,540){\circle{8}}} \Cgre{\put(68,514){\circle{8}}  \put(131,514){\circle{8}}
\put(197,514){\circle{8}}} \Cred{\put(59,482){$\times$} \put(122,482){$\times$} \put(188,482){$\times$}}
\Cblu{\put(55,456){$\times$} \put(87,456){$\times$} \put(118,456){$\times$} \put(154,456){$\times$}
\put(186,456){$\times$} } \Cgre{\put(55,446){\circle{8}} \put(118,446){\circle{8}} \put(186,446){\circle{8}}}
\put(55,497){\line(0,1){6}} \put(185,497){\line(0,1){6}} \put(38,487){\small -T'} \put(190,487){\small T'}
\put(116,420){\line(1,0){6}} \put(116,580){\line(1,0){6}} \put(90,418){$-i\pi$} \put(100,578){$i\pi$}
\put(210,536){$\cdot$} \put(210,550){$\cdot$} \put(210,510){$\cdot$} \put(215,536){$i\pi/2$}
\put(215,550){$i\pi(1+a)/2$} \put(215,510){$i\pi a/2$} \put(111,554){\vector(-2,-1){21}}
\put(175,554){\vector(-2,-1){21}} \put(110,512){\vector(-2,-1){23}} \put(173,512){\vector(-2,-1){23}}
\put(62,487){\vector(2,1){23}} \put(127,487){\vector(2,1){23}} \put(62,448){\vector(2,1){21}}
\put(127,448){\vector(2,1){21}} \put(275,500){\vector(1,0){40}} \put(350,500){\vector(1,0){190}}
\put(440,410){\vector(0,1){200}} \put(522,590){$\theta$}
\put(517,588){\line(0,1){13}}\put(517,588){\line(1,0){13}} \Cblu{\put(380,540){\circle{8}}
\put(440,540){\circle{8}} \put(500,540){\circle{8}}  \put(376,456){$\times$} \put(436,456){$\times$}
\put(496,456){$\times$}}
 \put(380,497){\line(0,1){6}} \put(500,497){\line(0,1){6}}
\put(363,487){\small -T'} \put(505,487){\small T'} \put(437,420){\line(1,0){6}} \put(437,580){\line(1,0){6}}
\put(410,418){$-i\pi$} \put(420,578){$i\pi$}
\end{picture}
\caption{Roaming limit $\left(a\to i\frac{T}{2\pi}\right)$} \label{roam}
 \end{figure}

In fact, considering $a=m+i(2n+1)\frac{T}{2\pi}+\delta$, in the 
limit $\delta\to 0$ poles and zeroes move according to Figure \ref{roam}, 
leading to the Ising reflection amplitude (\ref{KI'ell}) with period $T'$.

 \vspace{3.5cm}

It is now clear the reason why we have to use a reflection amplitude with real period half of the $S$-matrix one.
In fact, contrary to the ordinary roaming limit in which $a=1+i C$ and $C\to\infty$, now the imaginary part of
$a$ is sent to a finite value $\frac{T}{2\pi}$. Hence, since in the reflection amplitude the parameter $a$ is
multiplied by a factor $\frac{1}{2}$ absent in the expression of the $S$-matrix, poles and zeroes are shifted on
the real axis by $\frac{T}{4}$ instead of $\frac{T}{2}$, and, in order to reach the same position, at the
beginning they have to be located at half the initial distance.

Figure \ref{roam} refers to the case $m=n=0$, but the same final situation will take place for any
$n\in\mathbb{Z}$ and for $m=0,3\,(\textrm{mod}\,4)$. On the other hand, in the case $m=1,2\,(\textrm{mod}\,4)$ we
need to consider an initial $K(\theta)$ with an Ising factor $K_{I}'(\theta)$ instead of $K_{I}''(\theta)$,
recovering at the end $K_{I}''(\theta)$.

\vspace{1.5cm}

In principle, it is possible to extend to the elliptic case also the general reflection amplitude (\ref{Ksinh}),
defining
\begin{equation}
K(\theta)=K_{I}''(\theta)f(\theta)G(\theta)\,,
\end{equation}
where
\begin{eqnarray}
G(\theta)&=&\frac{\textrm{sn}\left[\frac{\textbf{K}}{i\pi} \left(\theta+i\frac{\pi}{2}
(1+a)\right)\right]}{\textrm{sn}\left[\frac{\textbf{K}}{i\pi} \left(\theta-i\frac{\pi}{2} (1+a)\right)\right]}
\frac{\textrm{sn}\left[\frac{\textbf{K}}{i\pi} \left(\theta+i\frac{\pi}{2}
(2-a)\right)\right]}{\textrm{sn}\left[\frac{\textbf{K}}{i\pi} \left(\theta-i\frac{\pi}{2}
(2-a)\right)\right]}\nonumber\\
&\times&\frac{\textrm{sn}\left[\frac{\textbf{K}}{i\pi} \left(\theta-i\frac{\pi}{2}
(1-E)\right)\right]}{\textrm{sn}\left[\frac{\textbf{K}}{i\pi} \left(\theta+i\frac{\pi}{2} (1-E)\right)\right]}
\frac{\textrm{sn}\left[\frac{\textbf{K}}{i\pi} \left(\theta-i\frac{\pi}{2}
(1+E)\right)\right]}{\textrm{sn}\left[\frac{\textbf{K}}{i\pi} \left(\theta+i\frac{\pi}{2}
(1+E)\right)\right]}\nonumber\\
&\times&\frac{\textrm{sn}\left[\frac{\textbf{K}}{i\pi} \left(\theta-i\frac{\pi}{2}
(1-F)\right)\right]}{\textrm{sn}\left[\frac{\textbf{K}}{i\pi} \left(\theta+i\frac{\pi}{2} (1-F)\right)\right]}
\frac{\textrm{sn}\left[\frac{\textbf{K}}{i\pi} \left(\theta-i\frac{\pi}{2}
(1+F)\right)\right]}{\textrm{sn}\left[\frac{\textbf{K}}{i\pi} \left(\theta+i\frac{\pi}{2} (1+F)\right)\right]}
\end{eqnarray}
and
\begin{eqnarray}
f(\theta)&=&\left\{\frac{\vartheta_{4} \left[\frac{1}{2 i}\left(\theta+i\frac{\pi}{2}(1+a)\right)\right]}
{\vartheta_{4}\left[\frac{1}{2 i}\left(\theta-i\frac{\pi}{2}(1+a)\right)\right]}
\frac{\vartheta_{4}\left[\frac{1}{2 i}\left(\theta+i\frac{\pi}{2}(2-a)\right)\right]}
{\vartheta_{4}\left[\frac{1}{2 i}\left(\theta-i\frac{\pi}{2}(2-a)\right)\right]}\right\}^{2}
\nonumber\\
&\times&\left\{\frac{\vartheta_{4}\left[\frac{1}{2 i} \left(\theta-i\frac{\pi}{2}(1-E)\right)\right]}
{\vartheta_{4}\left[\frac{1}{2 i}\left(\theta+i\frac{\pi}{2}(1-E)\right)\right]}
\frac{\vartheta_{4}\left[\frac{1}{2 i}\left(\theta-i\frac{\pi}{2}(1+E)\right)\right]}
{\vartheta_{4}\left[\frac{1}{2 i}\left(\theta+i\frac{\pi}{2}(1+E)\right)\right]}
\right\}^{2}\nonumber\\
&\times&\left\{\frac{\vartheta_{4}\left[\frac{1}{2 i} \left(\theta-i\frac{\pi}{2}(1-F)\right)\right]}
{\vartheta_{4}\left[\frac{1}{2 i}\left(\theta+i\frac{\pi}{2}(1-F)\right)\right]}
\frac{\vartheta_{4}\left[\frac{1}{2 i}\left(\theta-i\frac{\pi}{2}(1+F)\right)\right]}
{\vartheta_{4}\left[\frac{1}{2 i}\left(\theta+i\frac{\pi}{2}(1+F)\right)\right]}\right\}^{2}
\end{eqnarray}

In order to avoid poles with real part in the physical strip, both the parameters $E$ and $F$ have to lie in the
interval $[0,1]$. Furthermore, the roaming limit leads to an Ising reflection factor only for certain specific
choices of $E$ and $F$ in terms of $a$.

\vspace{1cm}

\resection{Conclusions}

In this paper we have studied the physical effects induced by an infinite number of boundary resonance states in
certain integrable quantum field theories. The main result of our analysis regards the ultraviolet regime shown
by such kind of theories. We have studied, in particular, in the elliptic generalization of the Ising model, the
short--distance behaviour of the effective central charge and of the 
one--point function of the energy operator,
which in the ordinary case scale, respectively, as a constant and as $\frac{1}{r}$. In the elliptic case,
instead, both these quantities display oscillations as $r\to 0$, and do not have a definite limit. This leads to
the physical interpretation that the infinite resonance states with increasing mass present on the boundary never
decouple from the theory, even at very short distances, making meaningless in this case the concept of scaling
limit.

The other interesting feature emerged from this work is the crucial role played by the choice of the real
periodicity. In fact, various theories with different real periods for the scattering and the reflection
amplitudes lead to the same theory in the limit $\ell\to 0$, because the ordinary case corresponds to
$T\to\infty$. However, since for $\ell>0$ this period is finite, it is necessary to treat it carefully. Indeed,
the roaming limit procedure, which connects the Sinh-Gordon model to the Ising model, can be consistently
generalized to the elliptic case only imposing that the reflection amplitude has half real period with respect to
the $S$-matrix. This has a physical implication on the nature of the spectra of the theory: once the bulk
spectrum of resonance states is given, the boundary one is then automatically fixed.

\vspace{1cm}

\begin{flushleft}\large
\textbf{Acknowledgements}
\end{flushleft}

V.R. thanks SISSA for a pre-graduate grant and for hospitality.

\end{document}